\documentclass{pnastwo}

\usepackage{cite}  

\usepackage{graphicx}
\usepackage{PNAStwoF}
\usepackage{amssymb,amsfonts,amsmath}

\begin{document}

\title{Hydrodynamic capture of microswimmers into sphere-bound orbits}

\author{Daisuke Takagi\affil{1}{Applied Math Lab, Courant Institute, New York University, New York, New York 10012}\affil{2}{Department of Mathematics, University of Hawaii at Manoa, Honolulu, Hawaii 96822},
J\'{e}r\'{e}mie Palacci\affil{3}{Department of Physics, New York University, New York, New York 10003},
Adam~B.~Braunschweig\affil{4}{Department of Chemistry, University of Miami, Coral Gables, Florida 33146},
Michael~J.~Shelley\affil{1}{}
\and
Jun Zhang\affil{1}{}\affil{3}{}}

\contributor{Submitted to Proceedings of the National Academy of Sciences
of the United States of America}

\maketitle

\begin{article}
\begin{abstract}
Self-propelled particles can exhibit surprising non-equilibrium
  behaviors, and how they interact with obstacles or boundaries
  remains an important open problem.  Here we show that chemically
  propelled micro-rods can be captured, with little change in
  their speed, into close orbits around solid spheres resting on or
  near a horizontal plane. We show that this interaction between
  sphere and particle is short-range, occurring even for spheres
  smaller than the particle length, and for a variety of sphere
  materials. We consider a simple model, based on lubrication theory,
  of a force- and torque-free swimmer driven by a surface slip (the
  phoretic propulsion mechanism) and moving near a solid surface.
  The model demonstrates capture, or movement towards the surface, and
  yields speeds independent of distance. This study reveals the
  crucial aspects of activity-driven interactions of self-propelled
  particles with passive objects, and brings into question the use of
  colloidal tracers as probes of active matter.
  \end{abstract}



\section{Significance}
How autonomous microswimmers interact with their environment can be strikingly different than for passive particles, and often give rise to a wealth of non-equilibrium phenomena. We report surprising observations of synthetic swimmers being captured by, and then orbiting, spherical particles until they are kicked out by thermal fluctuations. This complex behavior can be explained in part by a simple model with the basic ingredients at work: fluid slipping on the swimmer surface and fluid coupling with nearby boundaries. Our study stresses outcomes of activity-driven interactions between swimmers and other objects that were overlooked in previous studies. Our findings reveal a richer range of possible interactions in active matter, and impact statistical physics, soft matter, fluid dynamics, and biophysics.\newline

\dropcap{B}iological systems demonstrate that the presence of obstacles, which could be a solid surface or a group of cells, profoundly affects the autonomous movement of motile microorganisms. Motile cells can aggregate~\cite{berke2008}
and move in circles on surfaces~\cite{lauga2006}, reverse directions
when they are spatially constricted~\cite{cisneros2006}, and migrate
preferentially through an array of V-shaped
funnels~\cite{galajda2007}. Bacteria may enhance the diffusivity of
surrounding tracer particles~\cite{wu2000, leptos2009}, drive ratchets
into rotary motion~\cite{dileonardo2010, sokolov2010}, and form large
rotating structures through collective
movements~\cite{schwarz2012}. Interactions between bacteria and the
physical environment are widely documented but complicated because, in
addition to possible collisional~\cite{kantsler2013} and
hydrodynamic~\cite{drescher2011} effects, there are unknown responses
associated with behavior.

Recent technological advances have enabled the fabrication of synthetic microswimmers that convert chemical energy
into directional motion~\cite{paxton2004, fournier2005, gibbs2009, ebbens2010,
  sengupta2012}. Their movements can collectively drive the system out
of equilibrium and exhibit large-scale phenomena such as
schooling~\cite{ibele2009} and clustering~\cite{theurkauff2012,
  palacci2013}. Studying the dynamics of self-propelled particles is
important because they may have useful ensemble properties, inspire
new designs for smart materials, and find many applications including
microfluidic mixing devices and cargo transport~\cite{pumera2011,
  wang2012, patra2013}.

Synthetic swimmers are arguably simpler than bacteria and offer
insight into the effects of obstacles and confinement on
self-propelled particles. Here we employ a widely studied system
consisting of gold-platinum (Au-Pt) segmented rods immersed in an
aqueous hydrogen peroxide solution
(H$_2$O$_2$)~\cite{paxton2004}. Previous studies propose that these
rods propel themselves \textit{via} self-electrophoresis, which
generates a slip flow along the rod surface
~\cite{wang2006,moran2011}. Recent studies show that synthetic
swimmers interact with rigid boundaries, say by sliding along
walls~\cite{volpe2011,thutupalli2011} and enhancing the diffusivity of tracer
particles~\cite{mino2011}, but the mechanisms of the interactions
remain unclear.

\begin{figure}
\centerline{
\includegraphics[width=0.45\textwidth]{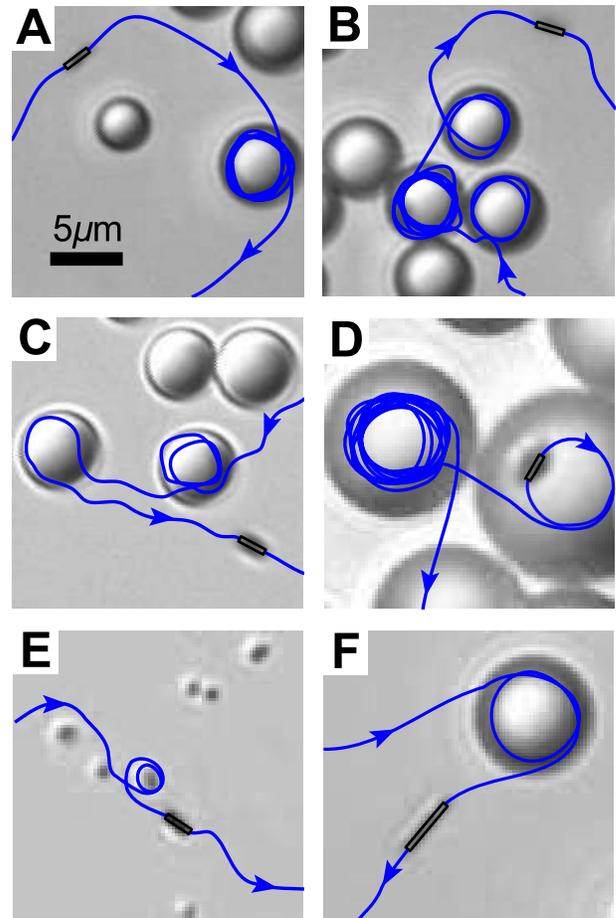}}
\caption{\label{fig:1} Sample trajectories of
  self-propelled rods orbiting around passive spheres. The spheres
  have diameter $\sim$6\,$\mu$m in (A-C), 11 and 14\,$\mu$m in (D),
  1\,$\mu$m in (E), and 9\,$\mu$m in (F), and are made of various
  materials (see main text). The rod length is 2\,$\mu$m in (A-E) and
  4\,$\mu$m in (F). The speed of the rods is on the order of
  20\,$\mu$m/s.}
\end{figure}

Here we show that self-propelled Au-Pt rods are captured and orbit
closely, with little decrease in their speed, around solid passive
spheres resting on a solid substrate (Fig.~1). We show that this
interaction between rod and sphere is short range, occurs for
spheres of various materials and sizes, including spheres below the
rod length. While the spheres appear to attract the rods, the
Stokesian fluid environment precludes any net force or torque being
exerted upon them. We explain some of these observations using a
simple model, based on lubrication theory, of a swimming particle
moving near a wall and propelled by a surface slip. This is a suitable
assumption for modeling the motion of phoretic swimmers and motile
ciliates~\cite{stone1996}.

\section{Results and Discussions}

\subsection{Experimental system} We fabricate Au-Pt rods with length $L=2\pm0.2\,\mu$m and diameter
$0.39\pm 0.04$\,$\mu$m following the method of electrochemical
deposition in anodic aluminum oxide
membranes~\cite{martin1999,banholzer2009}. They are immersed in a
H$_2$O$_2$ solution with typical concentration 15\% containing passive
spheres, with diameters of 1 to $\sim$100$\mu$m. Due to gravity, both
rods and spheres remain close to, and move along, the plane of the
microscope slide. The positions of the rods and spheres are tracked
using optical microscopy (Nikon Eclipse 80i, 40$\times$), a digital
camera (Lumenera Infinity 1-3), and image analysis (ImageJ and
Matlab).  In the absence of passive spheres, the moving rods turn, flip, and disperse over time as reported previously~\cite{takagi2013}.

\subsection{Capture of self-propelled rods by passive spheres} The presence of passive spheres significantly alters the trajectories
of self-propelled rods (Fig.~\ref{fig:1} and Movie~S1). When a
self-propelled rod encounters a sphere, it typically orbits around it
(Fig.~\ref{fig:1}A). The rod can move around a succession of spheres
and either continue to turn in the same sense (Fig.~\ref{fig:1}B) or
switch the handedness of its circular orbit (Fig.~\ref{fig:1}C).

This phenomenon of rods orbiting around spheres occurs across a wide
range of materials and sizes (Fig.~\ref{fig:1}D-F). We have used
rods of different lengths (1, 2, 4\,$\mu$m) and spheres of various
diameters (1, 3, 6, 9, 11, 14, 20, 125$\,\mu$m) made of glass,
polymerized 3-(trimethoxysilyl)propyl methacrylate
(TPM)~\cite{sacanna2010}, poly(methyl methacrylate) (PMMA), and
polystyrene. These materials have different surface properties but this
does not affect the capture effect. Self-propelled rods
are captured by all spheres that remain close to the substrate by
gravity, including glass beads as small as 1$\mu$m in diameter (half
the rod length in Fig.~\ref{fig:1}E; see Movie~S2). On a sheet of
mica with cleavages which act as nearly vertical walls on a horizontal
surface, the rods are transitorily trapped along the walls in a
similar fashion to what has been reported
before~\cite{volpe2011}. This demonstrates that the rods are captured
by confining walls with a variety of shapes and materials, and
suggests that the capture effect may be hydrodynamic in origin.

\begin{figure}
\centerline{\includegraphics[width=0.4\textwidth]{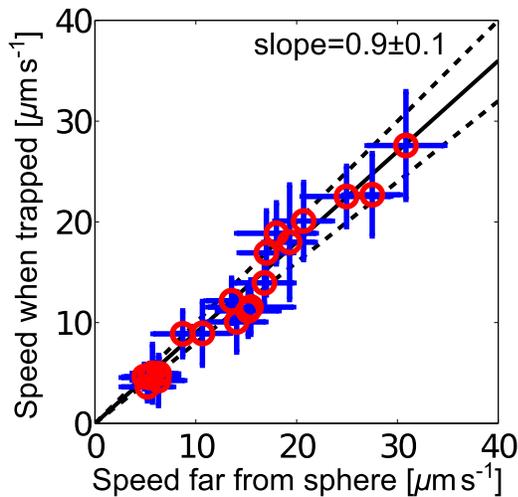}}
\caption{\label{fig:2} While self-propelled rods are
  trapped, they move at speeds comparable to when they are far away
  from spheres. Symbols and error bars respectively show the temporal
  mean and standard deviation in speed of each rod. Solid and dashed
  lines have slopes $0.9\pm0.1$. Data collected at different H$_2$O$_2$
  concentrations and over many events.}
  \end{figure}

\subsection{Capture phenomenon with no additional drag} Self-propelled rods nearly maintain their typical speed while they
orbit around spheres (Fig. \ref{fig:2}A). While a rod is
captured, it remains in a narrow fluid region between the sphere and
the horizontal glass substrate, and shows little change in speed
in this confined space. 

\begin{figure}
\centerline{\includegraphics[width=0.5\textwidth]{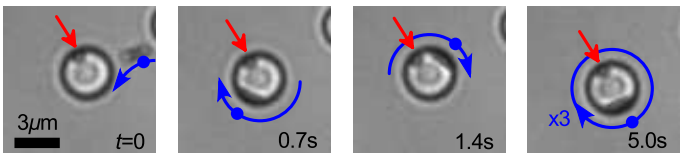}}
\caption{\label{fig:3} Image sequence of a solid sphere of diameter $\sim 3\mu$m which hardly rotates while a 2\,$\mu$m-long self-propelled rod orbits around it. Red arrows point to a marker on the sphere. Blue dots and arrows show the location and direction of motion of the orbiting rod. }
\end{figure}

Do spheres also rotate while a rod orbits around them? To answer this
we use inert TPM spheres with an embedded piece of hematite acting as
a marker~\cite{sacanna2012}. The rotational diffusion of the sphere is
slow compared to the orbiting time of the rod, and the orientation of
the marker remains apparently unchanged (Fig.~\ref{fig:3} and Movie~S3). A Fourier frequency analysis of sphere displacements shows that
while the sphere does fluctuate slightly because of Brownian motion,
there is also an oscillation of the sphere at the same frequency as
the rod orbital motion. However, this effect is quite small.

The nearly constant rod speed and the lack of sphere rotation suggest
that the rods hardly experience any additional drag despite moving
close to the solid spheres. This direct experimental evidence supports
the hypothesis that the rods generate a local propulsive flow along
their surface ~\cite{wang2006, moran2011} as examined in our theory
below.

\subsection{Tentative scenarios of capture} There are several possible mechanisms of trapping. Chemical gradients
induced by fuel-consuming rods may affect rod movement. However, in
our system diffusion quickly restores any local changes in fuel level
because the Peclet number $Pe=UL/D\sim 0.1$ is small, where $U\sim
20\,\mu$m/s is the typical speed, $L\sim 6\,\mu$m the sphere diameter,
and $D\sim 10^3\,\mu$m$^2$/s the fuel diffusivity. The consumption rate of the fuel is negligible over the typical duration of our experiments. This is consistent
with the swimmers having little loss in speed when orbiting a
sphere. Long-range hydrodynamic effects may enable swimmers to
approach a solid surface if they are represented by an extensile force
dipole (pusher)~\cite{berke2008, spagnolie2012}, though our rods
experience only short-range interactions with spheres as confirmed
below. The rods are slightly curved and tend to move in curved paths
with typical curvature $\kappa\sim
0.12\,\mu$m$^{-1}$~\cite{takagi2013}, so they could slide along
flat~\cite{van2008} and curved surfaces with curvature much less than
$\kappa$~\cite{van2009}. However, we observe rods orbiting around
spheres with diameter as small as 1\,$\mu$m, which has curvature much
greater than $\kappa$.

In our interpretation, aside from thermal fluctuations, the rods move
towards a surface because the local flow field is modified by its
presence. This is demonstrated in a simplified model based on
lubrication theory. Such a motion would not occur for an object towed
by a force parallel to a solid boundary due to the symmetry and
time-reversibility of Stokes flow. The key ingredient in our model is
a prescribed slip on the swimming body, which captures a phoretic
propulsion mechanism.

\section{Short-range hydrodynamic capture}

\begin{figure}
\centerline{\includegraphics[width=0.5\textwidth]{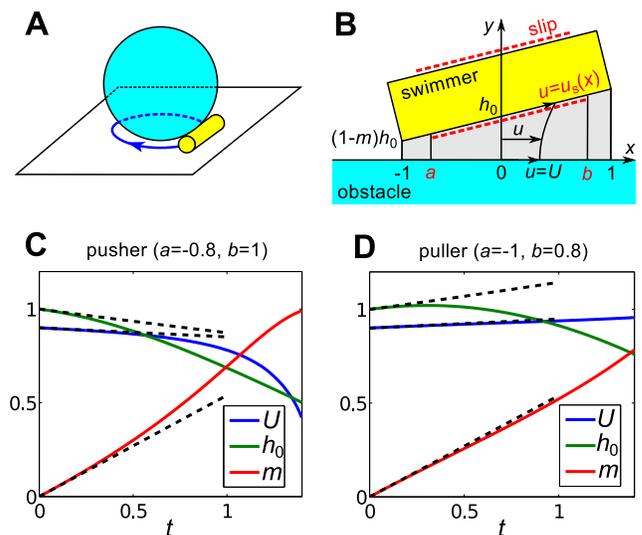}}
\caption{\label{fig:4} (A) Sketch of a rod orbiting
  around a sphere on a horizontal surface. (B) Theoretical
  configuration of a swimming body near a rigid wall in the reference
  frame of the swimmer. The wall approximates the local surface of the
  sphere and translates to the right. Dashed lines show the region of
  fluid slip. (C,D) Predictions for the swimming speed $U$, distance
  $h_0$ away from the wall, and re-scaled slope $m$ of a swimmer with
  fluid slipping over (C) $-0.8<x<1$ and (D) $-1<x<0.8$. Dashed lines
  show the early-time regime (Eq.~\ref{x0}-\ref{m}). }
\end{figure}

\subsection{Mathematical model} While our experimental system is three dimensional with curved
boundaries (Fig.~\ref{fig:4}A), for simplicity we consider a
two-dimensional swimmer moving above a flat immobile wall
(Fig.~\ref{fig:4}B). Thermal fluctuations are neglected. We adopt a
Cartesian coordinate system $(x,y)$ in the reference frame that
translates with the swimmer along the wall, where $x$ is scaled by a
half of the length of the swimming body $L/2$ and $y$ is scaled by a
typical distance $H$ between the swimmer and the wall. The wall is
represented by $y=0$ and translates to the right with speed $U$, where
all speeds are scaled by a typical slip velocity $V$ on the
swimmer. The wall-facing surface of the swimmer is represented by
$y=h$ with $h=h_0(1+mx)$, $-1\leq x \leq 1$, where $h_0$ is a
re-scaled distance from the wall to the center of the swimmer and $m$
is a re-scaled slope of the swimmer relative to the wall. Note that
$m=1$ corresponds to the front end of the swimmer (the left end in
Fig.~\ref{fig:4}B) hitting the wall.  Our aim is to predict how $U$,
$h_0$, and $m$ evolve over time, given a prescribed slip speed
$u_s(x)$ on the surface of the swimmer.

Lubrication theory describes the fluid flow between the swimmer and
the wall, provided that both the Reynolds number of the flow and
$\delta=H/L$ are small. Our approach is similar to that used in
modeling of crawling snails~\cite{chan2005}. The flow is primarily in
the $x$ direction and governed by
$d^2 u/dy^2={dp}/{dx}$, where $u$ is the flow speed, $p$ is the pressure scaled by $\mu V
H/L^2$, and $\mu$ is the dynamic viscosity of the fluid. The solution is given by
\begin{equation}
\label{speed}
u=\frac{1}{2}\frac{dp}{dx}y(y-h)+\frac{y}{h}(u_s-U)+U,
\end{equation}
which satisfies the boundary conditions $u=U$ on $y=0$ and $u=u_s(x)$ on
$y=h$. Substituting Eq.~(\ref{speed}) into the condition $\partial h/\partial t+(\partial/\partial x)\int_0^h u\,dy=0$ that the mass is conserved, we obtain
\begin{equation}
\label{dpdx}
\frac{dp}{dx}=\frac{6}{h^3}\left[\dot{h}_0 x(2+mx)+\dot{m}h_0 x^2+h(U+u_s)-2q_0\right].
\end{equation}
Here $q_0$ is a constant, which is obtained by integrating Eq.~(\ref{dpdx}) with respect to $x$ and requiring that $p=0$ at the ends $x=\pm 1$. Eqs.~(\ref{speed}) and (\ref{dpdx}) are used to convert the conditions that the swimmer is force- and torque-free,
\begin{equation}
\int_{-1}^{1} \frac{du}{dy}\bigg|_{y=h}\,dx=\int_{-1}^1 p \,dx=\int_{-1}^1 xp \,dx=0,
\end{equation}
into a system of ordinary differential equations. One consequence is an expression for the
swimming speed
\begin{equation}
\label{mikesdiscovery}
U=\frac{m}{\ln\left(\frac{1+m}{1-m}\right)} \int_{-1}^{1}\frac{u_s(x)\,dx}{1+mx},
\end{equation}
which is independent of the distance $h_0$ between the swimmer and the
wall. In addition, no force is exerted on the
wall. These are consistent with the rods nearly maintaining their
speed and exerting little drag on the spheres in our experiments
(Fig.~\ref{fig:2},\ref{fig:3}).

We now consider a simple class of swimmers with constant slip $u_s=1$
on a portion of their surface $a<x<b$ and no slip $u_s=0$
elsewhere. This assumption is motivated by the previous
hypothesis that the rods generate a local flow on their surface. At
early times, a swimmer that is initially parallel to the wall with
$h_0=1$ evolves according to the asymptotic solutions
\begin{eqnarray}
U &\sim& \frac{1}{2}(b-a)-\frac{15}{32}(b^2-a^2)(b-a-(b^3-a^3))t,\label{x0}\\
h_0 &\sim& 1- \frac{3}{8}(b^2-a^2)t,\label{h0}\\
m &\sim& \frac{15}{8}(b-a-(b^3-a^3))t,\label{m}
\end{eqnarray}
valid to second order in time. Eq.~(\ref{m}) shows that,
regardless of the exact region of slip, all swimmers with fluid
slipping on a majority of their surface orient towards the
wall. Temporal changes in $m$, $h_0$, and $U$ are shown for two
typical swimmers with slip everywhere except either near its front or
back end of the swimmer (Fig.~\ref{fig:4}C,D). In both cases, the
swimmer approaches the wall, and its changes in speed can be related
to the relative position of its no-slip region to the wall.  This
simple model suggests that self-propelled particles can be captured by
solid surfaces without applying a mean force or torque.

\subsection{Consequences of the model} In our experiments there are effectively two walls: the spherical
surface and the underlying substrate. Our theory would apply not only
to the sphere as the wall but also to the substrate. Lubrication
effects may induce self-propelled rods to move vertically downwards in
the same direction as the gravitational force, which likely reduces
the sedimentation height of self-propelled rods compared to that of
diffusive rods~\cite{takagi2013}. As a result the self-propelled rods
orient towards the thin gap between the walls of the sphere and the
substrate, though the effect may be weaker in 3D than in our 2D model.

\begin{figure}
\centerline{\includegraphics[width=0.5\textwidth]{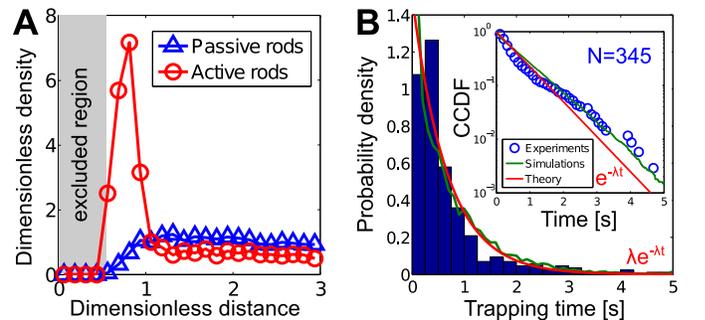}}
\caption{\label{fig:5} (A) Experimental data of the
  effective density of a suspension of rods, scaled by the expectation
  if the rods were distributed uniformly. This is plotted against the
  distance from the lowest point of a sphere in units of the sphere
  radius, based on over 4000 measurements of the distance over
  time. Below the sphere, the density of passive rods falls while the
  density of self-propelled rods peaks. The excluded region is where
  the rods are physically unable to enter. (B) Normalized histogram of
  trapping times of self-propelled rods orbiting around spheres. Inset
  shows the complementary cumulative distribution function (ccdf) denoting the probability of rods remaining trapped beyond a
  specified time. Red curves are exponential distributions with rate
  parameter $\lambda=1.5$\,s$^{-1}$. Green curves are simulated trapping times (see Materials and Methods).}
\end{figure}

To verify experimentally that only short-range interactions occur
between self-propelled rods and passive spheres, we consider the
distribution of rod positions relative to the center of a nearby
sphere (Fig.~\ref{fig:5}A). This is measured by sampling the
distance from a given rod to the nearest sphere over time, excluding
cases where the rod has multiple spheres within three sphere
radii. For passive rods in water, the density is nearly uniform
everywhere except below the sphere, where the density decreases to
zero. Active rods also have a uniform distribution away from the
sphere, but in contrast to passive rods now show a sharp peak below
the sphere. These features for self-propelled rods are consistent with
the hypothesis of capture through short-range interactions. Further,
hydrodynamic interactions induced by the rod's propulsive mechanism
enable the spheres to capture self-propelled rods but not passive
ones.

\subsection{Capture time statistics} While our deterministic
model predicts that swimmers move closer to a solid surface,
fluctuations may cause them to move away or remain at some
characteristic distance. In our experiments the trapping is transitory, and we observe that the
trapping times are typically less than 5 seconds and have an
exponential distribution with decay rate $\lambda\sim1.5$\,s$^{-1}$
(Fig.~\ref{fig:5}B). The trapping time-scale ($\lambda^{-1}$) is
comparable to the time-scales of rotational diffusion and duration
between stochastic flips ($\sim 1$s)~\cite{takagi2013} . 

In rare instances, the rods remain trapped on the order of
minutes. Highly curved rods that move in small circles tend to either
remain trapped for relatively long times or are hardly trapped,
depending on whether they approach a sphere with the trajectory
curving towards or away from the sphere, respectively. One possible
interpretation of this is that the rods are curved towards the sphere
while they are trapped until thermal fluctuations flip them over
\cite{takagi2013} and enable them to escape. 

We have developed a numerical model by including thermal
fluctuations in the orientational dynamics and defining a critical
angle for escape (see Materials and Methods). This yields a Poisson process for escape times, and
reproduces the exponential distribution found in Fig.~\ref{fig:5}B. The
relative residence times of trapped {\it vs} untrapped particles would
then yield the relative densities in Fig.~\ref{fig:5}A. While thermal fluctuations interacting with the capture effect seem to
explain the basic mechanisms at work, a more predictive theory should
include swimmer geometry and detailed chemistry.

\section{Conclusion}
Suspensions of self-propelled and passive particles reveal new aspects
of dynamics in active matter. Swimmers with an effective slip on a
majority of their surface, including phoretic particles~\cite{wang2006, moran2011} and motile ciliates~\cite{stone1996},
move without causing much disturbance to colloidal particles. This
suggests that phoretic particles can move stealthily, with little
perturbation to the surrounding fluid and objects within it. Still,
the swimmers themselves are captured by solid boundaries and can
accumulate in spatially constricted regions, unlike passive
particles. This may offer insight into the hydrodynamic effects in
cell adhesion and feeding, as well as suggest strategies to filter
groups of swimmers through the confined spaces of capillaries,
microfluidic devices, and porous media. Our results also bring into
question the use of colloidal tracers as probes in out-of-equilibrium
systems, as for example in measuring the effective temperature of an
active bath \cite{wu2000,leptos2009,mino2011}. The tracers may not
fully capture the complex swimmer movements and enhanced fluid mixing
which could emerge in concentrated suspensions. We speculate that
adding a payload to phoretic swimmers such as studied here would
markedly affect how they interact with the surrounding fluid and with
obstacles.

\begin{acknowledgments}
We thank S. Sacanna for the hematite-encapsulating spheres
used in this study. We acknowledge support from NSF (NYU
MRSEC DMR-0820341, MRI-0821520, DMR-0923251, DMS-0920930), DOE
(DE-FG02-88ER25053), and AFOSR (FA-9550-11-1-0032).
\end{acknowledgments}

\begin{materials}
\section{Simulation of trapping times} To study the effect of thermal fluctuations on the trapping
  time, we carried out simple simulations of our model with the added
  effect of rotational Brownian motion. Consider a swimmer with fluid slip over 90$\%$ of
  its body length $L=2\,\mu$m ($a=-0.9$, $b=0.9$) moving at speed
  $U=20\,\mu$m/s at a typical distance H$\sim$0.2\,$\mu$m from the wall. We simulate the orientation angle $\theta$ of 10000
  swimmers which are initially parallel to a wall. The angle evolves
  according to the Langevin Equation
  \begin{equation}
  \theta(t+\Delta t)=\theta(t)+\Omega\Delta t +\sqrt{2 D_r \Delta t}X,
  \end{equation}
  where $\Omega\sim 3\,$/s is the typical rotation rate estimated using Eq.~\ref{m}, $D_r=1\,$/s
  is the rotational diffusion constant, and $X$ is a random variable
  with normal distribution. This shows that the trapping time, defined
  as the time it takes for the angle to decrease to a threshold value
  $-\pi/9$, is exponentially distributed in agreement with our experimental data (Fig.~\ref{fig:5}B).
  \end{materials}


\end{article}

\end{document}